\documentclass[conference]{IEEEtran}
\IEEEoverridecommandlockouts

\usepackage{cite}
\usepackage{amsmath,amssymb,amsfonts}
\usepackage{algorithmic}
\usepackage{graphicx}
\usepackage{textcomp}
\usepackage{xcolor}
\def\BibTeX{{\rm B\kern-.05em{\sc i\kern-.025em b}\kern-.08em
    T\kern-.1667em\lower.7ex\hbox{E}\kern-.125emX}}

\usepackage{booktabs,array,colortbl,tabularx}
\usepackage{pifont}
\usepackage[bookmarks=false]{hyperref}
\usepackage{booktabs}
\usepackage{microtype}
\usepackage{xspace}
\usepackage{enumitem}
\usepackage{balance}
\usepackage{stfloats}
\usepackage{colortbl}
\usepackage{multicol}
\usepackage{multirow}
\usepackage{pifont}
\usepackage{algorithm}

\usepackage[table]{xcolor}
\definecolor{oursbg}{HTML}{EDF3F8}

\begin{document}

\title{2D or 3D? Rethinking Token Pruning in 3D VLAs with Cross-Modal Correlation Awareness}

\author{%
  \IEEEauthorblockN{%
    Zihao Zheng\textsuperscript{1,*},
    Zhihao Mao\textsuperscript{1,*},
    Quanyou Li\textsuperscript{1},
    Hangyu Cao\textsuperscript{2},
    Sicheng Tian\textsuperscript{3},
    Lingyue Zhang\textsuperscript{4},\\
    Chenyue Li\textsuperscript{1},
    Ziyun Zhang\textsuperscript{1},
    Hong Gao\textsuperscript{5},
    Yuchen Huang\textsuperscript{5},
    Yutong Xu\textsuperscript{5},
    Guojie Luo\textsuperscript{1},
    Xiang Chen\textsuperscript{1,\ensuremath{\dagger}}%
  }%
  \IEEEauthorblockA{%
    \textsuperscript{1}\,School of Computer Science, Peking University\\
    \textsuperscript{2}\,School of Software and Microelectronics, Peking University\\
    \textsuperscript{3}\,School of Artificial Intelligence, Beijing Normal University\\
    \textsuperscript{4}\,School of Electronics Engineering and Computer Science, Peking University
    \quad \textsuperscript{5}\,ZTE Corporation%
  }%
  \thanks{\textsuperscript{*}\,Equal contribution.}%
  \thanks{\textsuperscript{\ensuremath{\dagger}}\,Corresponding author:
    Xiang Chen (\texttt{xiang.chen@pku.edu.cn}).}%
}

\maketitle

\begin{abstract}
3D vision-language-action (VLA) models process both image and point cloud tokens at each control step, increasing the cost of action prediction.
Token pruning can reduce this cost, but a token's contribution depends on which other tokens remain.
We present \textit{2Dor3D}, a framework that jointly selects which 2D and 3D tokens to retain under a fixed token budget.
Our analysis shows that either modality can compensate for the other within a patch, that neighboring correlations help explain differences in pruning risk, and that shared tokens couple pruning decisions across patches.
In one box-closing observation, two patches with nearly identical interaction scores differed eightfold in the action deviation caused by removing their image token.
Guided by these findings, \textit{2Dor3D} transfers cross-modal prediction residuals to retained tokens, uses neighborhood correlation entropy to explore more retention choices at local outliers, and updates pruning costs after each proposed exchange between coupled regions.
A fused GPU kernel packs the selected tokens into a shorter sequence for subsequent Transformer layers.
Across RLBench and LIBERO, \textit{2Dor3D} speeds up end-to-end inference by $2.06$--$2.10\times$ on MLA and $1.86$--$1.91\times$ on FiS-VLA, with task success rates 0.5--0.6 percentage points below the corresponding unpruned models.
\end{abstract}


\section{\textbf{Introduction}}
\label{sec:introduction}

Vision-language-action (VLA) models map visual observations and language instructions to robot actions and have become an important paradigm for embodied intelligence~\cite{rt2, openvla}.
Despite this progress, early VLA models relied primarily on 2D visual inputs~\cite{rt1} and still faced challenges in fine-grained manipulation~\cite{3dvla}.
Therefore, recent studies have explored 3D VLA models that jointly encode 2D images and 3D observations to improve spatial perception~\cite{geovla,spatialvla}. 

In 3D VLAs, 2D image tokens encode visual appearance and semantics, while 3D point cloud tokens capture geometric and spatial information~\cite{3dsvla,geovla,fisvla}. 
Models such as MLA jointly process both modalities through shared attention~\cite{mla}. 
Since both modalities describe the same physical scene, their representations exhibit both information overlap and modality-specific characteristics~\cite{imvotenet}, resulting in complex cross-modal correlations that reflect both redundancy and complementarity.
However, joint reasoning over both modalities incurs substantial computational and memory overhead~\cite{pointvla}.

Token pruning reduces inference cost by shortening the visual token sequence processed by subsequent layers~\cite{dynamicvit}.
For 3D VLA models, identifying redundant tokens requires accounting for the information overlap and complementarity between 2D and 3D modalities.
Related studies in 3D and multimodal vision assess token importance to guide pruning: Fast3D predicts global attention to estimate token importance~\cite{fast3d}; Geo3DPruner uses geometric correspondence to reduce cross-view redundancy~\cite{geo3dpruner}; TokenFusion replaces uninformative tokens with features from other modalities for cross-modal fusion~\cite{tokenfusion}.

These studies demonstrate the effectiveness of using token importance to guide pruning and fusion.
However, existing methods do not fully capture cross-modal information sharing and complementarity or inter-patch dependencies between 2D and 3D tokens, which we refer to as token correlations.
Characterizing these correlations is key to distinguishing replaceable redundancy from irreplaceable features and guiding 2D/3D token pruning, but presents three challenges:
\textit{Challenge \ding{172}:}
Within a patch, removing either the 2D or the 3D tokens alone may have little effect on the predicted action.
These individual pruning costs do not reveal whether both modalities can be removed together or whether at least one must remain.
\textit{Challenge \ding{173}:}
Even when two patches have similar cross-modal correlation scores, removing the same modality can affect their predicted actions differently.
Assessing the pruning risk of a patch therefore requires understanding how its correlations are distributed across the surrounding neighborhood.
\textit{Challenge \ding{174}:}
Shared tokens couple correlations across modalities and spatial patches. Pruning one token can therefore alter other correlations, making the joint impact of multiple pruning decisions difficult to assess and the global retention configuration difficult to determine.

To address these challenges, we first analyze inner-patch and inter-patch 2D/3D correlations and then examine how these correlations are coupled across patches.
\textit{Analysis \ding{172}:} Across the four retention states of a patch, changes in action-related channels reveal how the two modalities interact and how much of the full response either can preserve on its own.
We quantify these effects with Interaction Correlation and Conditional Correlation, respectively.
\textit{Analysis \ding{173}:} Two patches with nearly identical interaction scores can incur different pruning costs, motivating us to characterize their neighborhoods using Inter-Patch Entropy computed separately for interaction and conditional scores.
\textit{Analysis \ding{174}:} Pruning one token can alter correlations among the remaining tokens, so we extend the two metrics to these coupled relations to assess how one removal changes the cost of the next.


Based on these insights, we propose \textit{2Dor3D}, a framework that jointly selects which image and point cloud tokens to retain under a fixed token budget.
First, \textbf{Correlation-Aware Candidate State Modeling} estimates how much action-related information each patch retains with both modalities, 2D only, 3D only, or neither.
When only one modality is retained, we predict the features of the other modality from it and transfer the prediction residual to the retained tokens.
Second, \textbf{Correlation-Guided Pruning Search} uses neighborhood correlation scores and their entropy to explore more retention choices at local outliers.
It discards candidates that cannot meet the token budget and ranks the rest by estimated information loss.
Third, \textbf{Coupling-Aware Pruning Optimization} proposes exchanges between coupled regions and updates the affected pruning costs after each exchange to guide the next one.

Our contributions are summarized as follows:
\begin{itemize}
    \item We systematically analyze the conditionality, context dependence, and coupling of 2D/3D token correlations, revealing their implications for action prediction and joint pruning.
    \item We propose \textit{2Dor3D}, which combines correlation modeling, guided search, and coupling-aware optimization to automatically determine 2D/3D token selection configurations under a budget.
    \item We evaluate \textit{2Dor3D} on MLA and FiS-VLA using task success rate and end-to-end inference speedup in simulation, and task success rate and completion time on real robots.
\end{itemize}

\begin{figure*}[t]
    \centering
    \includegraphics[width=7in]{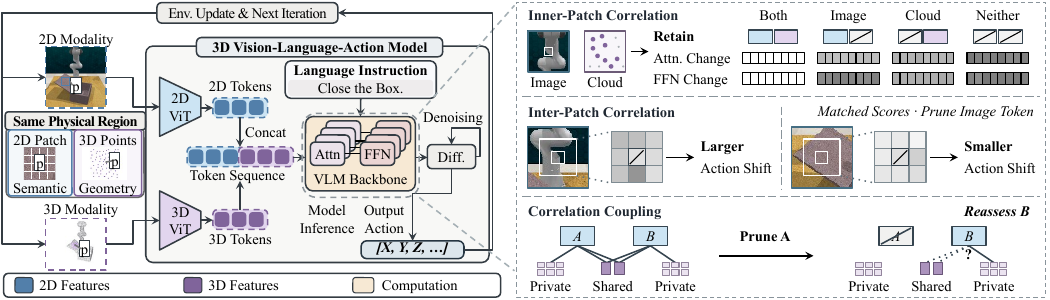}
    \vspace{-2mm}
    \caption{Challenges of Joint 2D/3D Token Pruning in 3D VLAs.}
    \label{fig:Challenges}
    \vspace{-4mm}
\end{figure*}

\section{\textbf{Background and Motivation}}
\label{sec:background}

\subsection{\textbf{3D Vision-Language-Action Models}}
\label{sec:background_1}

\subsubsection{\textbf{Encoding Paradigm and Model Architecture}}
3D vision-language-action (VLA) models use images, 3D observations, and language instructions to predict robot actions~\cite{leo,diffactor}.
MLA~\cite{mla} feeds image patches and tokenized 3D point groups directly into its language backbone. FiS-VLA~\cite{fisvla} first maps point groups into tokens with a lightweight 3D tokenizer. Its shared Vision Transformer (ViT) encodes both image and point tokens before passing their features to the shallow fast action module. Thus, visual sequence length affects computation inside MLA's language backbone and during feature extraction in FiS-VLA.

\subsubsection{\textbf{Cross-Modal Redundancy in 3D VLA Models}}
Aligned image and point cloud tokens can capture the same object boundary~\cite{3dsvla,mla}.
In 3D detection, TokenFusion~\cite{tokenfusion} replaces low-importance point tokens with corresponding image features.
For grasping, however, PointVLA~\cite{pointvla} uses point cloud geometry to distinguish a real object from its photograph.

\subsection{\textbf{Token Pruning for Model Inference Optimization}}
\label{sec:background_2}

\subsubsection{\textbf{Token Pruning for VLA Models}}
Visual token pruning reduces inference cost by shortening the sequence passed to subsequent layers~\cite{lightvla,iaprune}.
EfficientVLA~\cite{efficientvla} selects tokens for their relevance to the task and the diversity of their features.
SpecPrune-VLA~\cite{specprune-vla} combines attention from previous and current steps to select tokens for each action, then prunes further using layer-specific importance scores.

\subsubsection{\textbf{Challenges of 3D VLA Models' Token Pruning}}
Effective 2D/3D token pruning must account for both inner- and inter-patch correlations. Within a patch, pruning one modality depends on whether its counterpart is retained; across patches, neighborhood correlations can lead to different pruning risks despite similar local scores. These dependencies are further coupled, as pruning one token can affect the removability of others under a fixed budget. Therefore, we analyze 2D/3D cross-modal relationships in 3D VLA to derive insights for pruning design.

\section{\textbf{2D/3D Modality Correlation Analysis \\ for 3D Vision-Language-Action Models}}
\label{sec:analysis}

We first examine how 2D and 3D tokens interact and compensate for each other within a patch, then how correlation scores vary across neighboring patches.
Finally, we examine how pruning one token affects other 2D/3D correlations and the cost of subsequent pruning decisions.

\subsection{\textbf{Analysis of Inner-Patch 2D/3D Modality Correlation}}

To examine how the contribution of one modality depends on the other, we test all 256 patches of one box-closing observation with MLA~\cite{mla}. With the input and decoding noise fixed, we remove the selected 2D token, the eight 3D tokens nearest its spatial anchor, or both. We measure the Euclidean distance from the unpruned action in normalized action space. At patch~55, removing 2D or 3D alone yields deviations of 0.00225 and 0.00164, respectively. Removing both raises the deviation to 0.00457. At patch~243, removing both yields 0.00172, less than either individual removal.

\begin{figure}[t]
    \centering
    \includegraphics[width=3.3in]{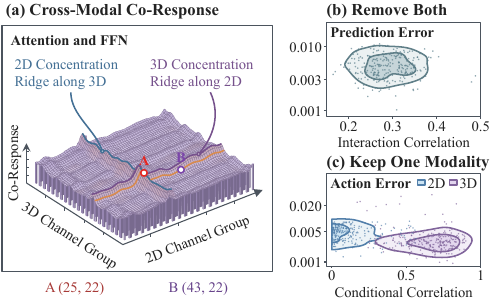}
    \vspace{-2mm}
    \caption{Analysis of 2D/3D Token Correlations within a Patch.}
    \label{fig:analysis1}
    \vspace{-4mm}
\end{figure}

For patch $i$, the two bits in $s\in\{11,10,01,00\}$ indicate whether its 2D token and 3D group are retained, respectively. Fig.~\ref{fig:analysis1}(a) shows which channels respond to both modalities. Let $z_s$ be a module's output at the action query in MLA layer~8 and $d_s=z_s-z_{00}$. We compute each score for the attention and feed-forward outputs and denote their average by $\langle\cdot\rangle$. \textbf{Interaction Correlation} measures the normalized departure from additivity,
\begin{equation}
C_i^{\mathrm{int}}=\left\langle
\frac{\|d_{11}-d_{10}-d_{01}\|_2}
{\|d_{11}\|_2+\|d_{10}\|_2+\|d_{01}\|_2}\right\rangle.
\label{eq:interaction_correlation}
\end{equation}
\textbf{Conditional Correlation} measures how much keeping modality $m$ reduces the squared response error relative to removing both. With $s_{\mathrm{2D}}=10$, $s_{\mathrm{3D}}=01$, and $[x]_+=\max(0,x)$,
\begin{equation}
C_{i,m}^{\mathrm{cond}}=\left\langle
\left[1-\frac{\|d_{11}-d_{s_m}\|_2^2}{\|d_{11}\|_2^2}\right]_+
\right\rangle.
\label{eq:conditional_correlation}
\end{equation}
Both denominators are positive for all 256 patches, and both scores lie in $[0,1]$. Fig.~\ref{fig:analysis1}(b) compares $C_i^{\mathrm{int}}$ with the error of predicting joint removal from individual action changes. In Fig.~\ref{fig:analysis1}(c), conditional scores correlate weakly with action error across patches in this frame. Across the four LIBERO suites, we follow DyQ-VLA~\cite{dyq} and rank scores and action errors across policy queries within each trajectory. Median Spearman $\rho$ is $-0.923$ for 2D only and $-0.927$ for 3D only.

\noindent\textbf{Insight \ding{172}:} Either modality can compensate for the other, so small individual pruning costs can conceal the cost of removing both.

\subsection{\textbf{Analysis of Inter-Patch 2D/3D Modality Correlation}}

\begin{figure}[t]
    \centering
    \includegraphics[width=3.3in]{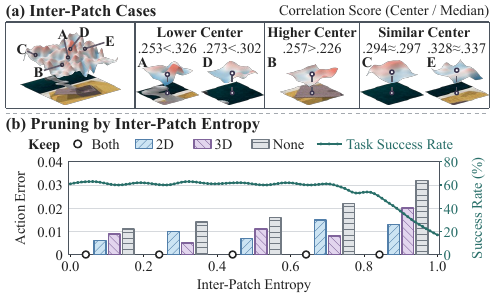}
    \vspace{-2mm}
    \caption{Analysis of Inter-Patch Entropy And Pruning Risk.}
    \label{fig:analysis2}
    \vspace{-4mm}
\end{figure}

We next compare patches with similar inner-patch scores but different neighborhood distributions.
In Fig.~\ref{fig:analysis2}~(a), patches~A and~B have nearly identical interaction scores (0.253 and 0.257). Removing the center 2D token nevertheless causes an eightfold difference in action deviation (0.02385 vs.\ 0.00293). Removing the 3D group yields a much smaller gap (0.00332 vs.\ 0.00258). The center score at~A falls below the median of its eight neighbors (0.326), whereas the score at~B exceeds its surrounding median (0.226). A pruning rule based on the center score alone would give the two 2D tokens similar priority.

We compute \textbf{Inter-Patch Entropy} separately for the interaction score and each directional conditional score over a $3\times3$ neighborhood $R$ clipped at the image boundary. When the neighborhood sum is positive, we express each patch's score as a share of that sum and divide the Shannon entropy of the shares by $\log |R|$. Low entropy means a few patches carry most of the score. Entropy is one for uniform positive scores and undefined for all-zero scores. The center-to-neighbor difference identifies which patches stand out locally. Fig.~\ref{fig:analysis2}(b) plots action error for each of the four retention states against inter-patch entropy, with task success rate on the right axis.

\noindent\textbf{Insight \ding{173}:} Combine the center scores with their spatial distributions when choosing which modality to prune.

\subsection{\textbf{Analysis of Complex Coupling Among 2D/3D Correlations}}

\begin{figure}[!b]
    \centering
    \vspace{-4mm}
    \includegraphics[width=3.3in]{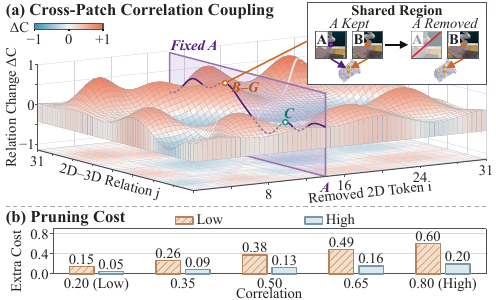}
    \vspace{-2mm}
    \caption{Analysis of Correlation Coupling across Patches.}
    \label{fig:analysis3}
\end{figure}

The preceding comparisons hold all other tokens fixed.
In Fig.~\ref{fig:analysis3}(a), patches~104 and~119, denoted by $A$ and $B$, share two of their eight 3D tokens in region $G$. Removing the 2D token at $A$ leaves $B$ and $G$ in the model. The purple slice traces the change in each surviving relation when $A$ is removed. The change is larger for $B$--$G$ than for $C$.

For two relations sharing $G$, hold $G$ fixed and let the two state bits indicate retention of their respective 2D endpoints. Equation~\eqref{eq:interaction_correlation} then gives their joint interaction, and Eq.~\eqref{eq:conditional_correlation} gives the response recovered by either surviving relation. In Fig.~\ref{fig:analysis3}(b), extra cost is the increase in the marginal cost of removing $B$ once $A$ is gone. At each interaction level, the bars compare this extra cost under low and high conditional compensation.

\noindent\textbf{Insight \ding{174}:} Correlation coupling can make individually inexpensive pruning choices costly together. After each removal, reassess candidates with strong interaction and little remaining compensation.

\section{\textbf{2D/3D Correlation-Aware Token Pruning}}
\label{sec:method}

\begin{figure*}[t]
    \centering
    \includegraphics[width=7in]{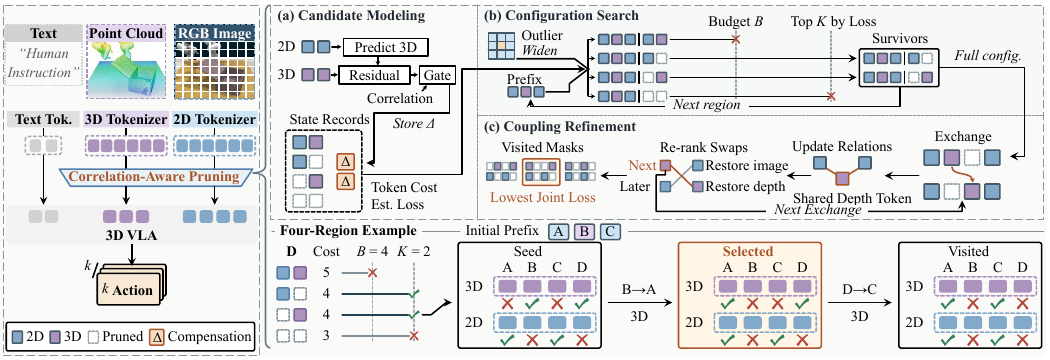}
    \vspace{-2mm}
    \caption{Overview of the \textit{2Dor3D} Framework for Joint 2D/3D Token Pruning.}
    \label{fig:method}
    \vspace{-4mm}
\end{figure*}

\subsection{\textbf{Correlation-Aware Candidate State Modeling}}
\label{sec:method_1}
The bits of $s_i=(u_i,v_i)$ specify patch $i$'s local requests to retain its 2D token and associated 3D tokens. Following Insight~\ding{172}, we estimate how much of the joint response each state preserves after compensation.

Offline, perturbations of held-out training observations with the backbone frozen provide the correlation targets from Sec.~\ref{sec:analysis}, and unpruned channel responses provide targets for the low-rank compensation maps. Online, a single backbone pass supplies regional attention and feed-forward features $h_i^{2D}$ and $h_i^{3D}$. For either modality $m$ and its counterpart $n$, let $r_i^n=h_i^n-F_{m\to n}h_i^m$ be the residual of a low-rank cross-modal prediction. A two-layer perceptron $\Phi_{\rm corr}$ processes both features and both residuals in a batch to estimate $\widehat C_i^{\rm int}$ and the two directional scores $\widehat C_{i,m}^{\rm cond}$. These estimates drive a linear-sigmoid gate $g_i^m$ for each direction.
Let $\alpha_i=\|h_i^{2D}\|_2+\|h_i^{3D}\|_2$ denote regional salience and $\eta_m\in(0,1)$ the recovery factor for modality $m$. The local losses for retaining both modalities, neither, or only $m$ (state $s_m$) are estimated as
\begin{equation}
\begin{aligned}
 \ell_i(11)&=0,\qquad \ell_i(00)=\alpha_i(1+\widehat C_i^{\rm int}),\\
 \ell_i(s_m)&=\alpha_i(1-\widehat C_{i,m}^{\rm cond})(1-\eta_m g_i^m).
\end{aligned}
\label{eq:local_loss}
\end{equation}
The interaction term penalizes departures from feature additivity in either direction. This conservative ranking can miss cases where joint removal causes less action deviation than a single removal. Once the mask is fixed, a patch retaining only $m$ adds the gated projection $g_i^m W_m r_i^n$ to its retained feature $h_i^m$ before compaction. A 2D-only patch draws its 3D residual only from removed tokens. If several patches compensate the same retained 3D token, we add the mean of their gated updates once.

\subsection{\textbf{Correlation-Guided Pruning Search}}
\label{sec:method_2}
We visit patches in row-major order. For each patch, we consider states $11$, $00$, and the feasible single-modality state with lowest local loss, if available. Following Insight~\ding{173}, we also test the other single-modality state when the center score falls below its neighbors' mean and the neighborhood entropy falls below the frame median, both for the same correlation metric.

We extend each surviving partial configuration with each candidate state of the current patch. Its score $\widehat{\mathcal L}(\mathbf s)=\sum_{i\in V}\ell_i(s_i)$ covers only the patches assigned so far, denoted by $V$. Let $T_i$ contain the 3D indices associated with patch $i$. A 3D token survives if any patch requests it. The physical token count is
\begin{equation}
 N(\mathbf s)=\sum_{i\in V} u_i+
 \left|\bigcup_{i\in V:\,v_i=1}T_i\right|.
 \label{eq:unique_count}
\end{equation}
If patch~104 requests neither modality and patch~119 requests 3D, their two shared 3D tokens remain in the sequence. Both patches drop their 2D tokens, so the coupling penalty in Eq.~\eqref{eq:joint_loss} can still apply.
We merge 256-bit 3D masks by bitwise OR and count the set bits. Since assigning more patches cannot reduce $N$, we discard candidates over budget $B$. Of the rest, we keep up to $K$ with the lowest partial scores. These candidates may leave different amounts of budget for later patches.

\subsection{\textbf{Coupling-Aware Pruning Optimization}}
Following Insight~\ding{174}, we connect patches $i<j$ with $T_i\cap T_j\ne\varnothing$. For each shared-token edge in $\mathcal E$, a correction adjusts the local loss estimate for ranking full configurations:
\begin{equation}
\begin{aligned}
 \widehat{\mathcal L}_{\rm joint}(\mathbf s)
   &=\widehat{\mathcal L}(\mathbf s)+\sum_{(i,j)\in\mathcal E}\Delta\psi_{ij}(\mathbf s),\\
 \Delta\psi_{ij}(\mathbf s)
   &=\lambda_{\rm cpl}\,a_i a_j\rho_{ij}\widehat C_{ij}^{\rm int}
     \bigl(1-\widehat C_{ij}^{\rm cond}(\mathbf s)\bigr).
\end{aligned}
\label{eq:joint_loss}
\end{equation}
Here $a_i=\mathbf 1[s_i\ne 11]$ flags a pruned patch, $\rho_{ij}=|T_i\cap T_j|/|T_i\cup T_j|$ measures overlap, and $\lambda_{\rm cpl}\ge0$ sets the penalty scale. The edge interaction estimate $\widehat C_{ij}^{\rm int}$ is calibrated offline from paired perturbations. After each exchange, we update the conditional estimate for surviving relations from cached features. Strong interaction and weak compensation increase the penalty.

After the scan, an exchange can release tokens from an earlier patch to restore tokens in a later one. From each surviving configuration, we test budget-feasible replacements and exchanges, updating the affected scores after each move. Each chain takes at most $L$ moves, may pass through a higher-loss state, and changes each patch at most once. We keep the feasible state with the lowest joint loss along the chain, retaining the starting configuration if no move improves it. Refinement requires no additional backbone pass.

\section{\textbf{Implementation}}
\label{sec:implementation}

\begin{table}[t]
    \centering
    \caption{Simulation with \textcolor{blue}{\textbf{MLA}}~\cite{mla} (2D/3D pruning ratios in \%)}
    \vspace{-2mm}
    \label{tab:main-mla}

    \footnotesize
    \setlength{\tabcolsep}{0.5pt}
    \renewcommand{\arraystretch}{0.98}
    \setlength{\aboverulesep}{0.2ex}
    \setlength{\belowrulesep}{0.2ex}
    \setlength{\cmidrulesep}{0.15ex}

    \resizebox{\columnwidth}{!}{%
    \begin{tabular*}{\columnwidth}{
        @{\extracolsep{\fill}}cc*{6}{c}@{}
    }
        \toprule
        \toprule
        \multirow{2}{*}{Type}
        & \multirow{2}{*}{Method}
        & \multicolumn{3}{c}{\textbf{RLBench}~\cite{rlbench}}
        & \multicolumn{3}{c}{\textbf{LIBERO}~\cite{libero}} \\
        \cmidrule(lr){3-5}
        \cmidrule(lr){6-8}
        &
        & SR$\uparrow$ (\%) & Speed$\uparrow$ & PR (\%)
        & SR$\uparrow$ (\%) & Speed$\uparrow$ & PR (\%) \\
        \midrule

        & Unpruned
        & \textbf{81.0} & 1.00x & 0/0
        & \textbf{90.0} & 1.00x & 0/0 \\
        \midrule

        \multirow{5}{*}{Att.}
        & FastV~\cite{fastv}
        & 78.5 & 1.74x & 75/75
        & 87.2 & 1.80x & 75/75 \\

        & SparseVLM~\cite{sparsevlm}
        & 78.5 & 1.75x & 75/75
        & 86.7 & 1.77x & 75/75 \\

        & SP-VLA~\cite{spvla}
        & 79.1 & 1.16x & 25/25
        & 87.1 & 1.14x & 25/25 \\

        & VLA-Pruner~\cite{vlapruner}
        & 79.7 & 1.69x & 75/75
        & 88.5 & 1.74x & 75/75 \\

        & Fast3D~\cite{fast3d}
        & 78.4 & 1.75x & 75/75
        & 87.3 & 1.71x & 75/75 \\
        \midrule

        \multirow{3}{*}{Fea.}
        & DivPrune~\cite{divprune}
        & 76.4 & 1.78x & 75/75
        & 86.0 & 1.78x & 75/75 \\

        & DART~\cite{dart}
        & 77.8 & 1.87x & 75/75
        & 87.8 & 1.89x & 75/75 \\

        & CDPruner~\cite{cdpruner}
        & 79.9 & 1.84x & 75/75
        & 88.7 & 1.89x & 75/75 \\
        \midrule

        Corr.
        & \textbf{\textit{2Dor3D}}
        & \textbf{80.5} & \textbf{2.06x} & \textbf{80/70}
        & \textbf{89.4} & \textbf{2.10x} & \textbf{80/70} \\
        \bottomrule
        \bottomrule
    \end{tabular*}%
    }
    \vspace{-4mm}
\end{table}

\begin{figure}[b]
    \centering
    \vspace{-4mm}
    \includegraphics[width=3.3in]{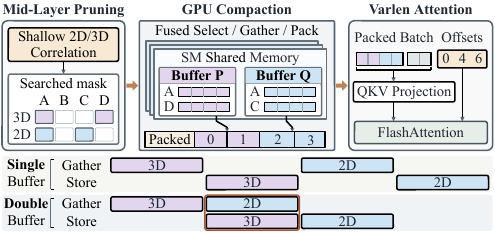}
    \vspace{-2mm}
    \caption{Implementation of the Proposed Token Pruning Framework}
    \label{fig:implementation}
\end{figure}

In Fig.~\ref{fig:implementation}, shallow features guide mask selection within the current backbone pass. In MLA, compaction at an intermediate language-model layer shortens the sequence for the remaining blocks. In FiS-VLA, we compact the tokens at an intermediate block of the shared vision encoder, after tokenization. The remaining encoder blocks and the fast action module process the shorter sequence. For a single-modality region, residual injection precedes compaction.
A fused GPU kernel then selects the retained rows, gathers their hidden states, and packs them contiguously in their original order. 
By eliminating the intermediate write and read, fusion saves an estimated 0.03\,ms per compaction on MLA. Most of the end-to-end speedup comes from processing fewer tokens in subsequent layers.

The kernel uses two shared-memory buffers, $P$ and $Q$, to overlap reads and writes across successive token blocks. 
While the current block is written from $P$ to global memory, the next block can be loaded into $Q$. 
Buffer $P$ can be reused once its write completes. 
Subsequent layers compute query, key, and value projections directly on the packed tokens. 
Variable-length FlashAttention~\cite{flashattention2} uses cumulative sequence lengths to identify sample boundaries and computes attention within each sample, avoiding padding to the longest sequence in the batch.

\section{\textbf{Experiments}}
\label{sec:experiments}

\subsection{\textbf{Setup}}
\label{sec:experiments_1}

\begin{table}[t]
    \centering
    \caption{Simulation with \textcolor{blue}{\textbf{FiS-VLA}}~\cite{fisvla} (2D/3D pruning ratios in \%)}
    \vspace{-2mm}
    \label{tab:main-fis}

    \footnotesize
    \setlength{\tabcolsep}{0.5pt}
    \renewcommand{\arraystretch}{0.98}
    \setlength{\aboverulesep}{0.2ex}
    \setlength{\belowrulesep}{0.2ex}
    \setlength{\cmidrulesep}{0.15ex}

    \resizebox{\columnwidth}{!}{%
    \begin{tabular*}{\columnwidth}{
        @{\extracolsep{\fill}}cc*{6}{c}@{}
    }
        \toprule
        \toprule
        \multirow{2}{*}{Type}
        & \multirow{2}{*}{Method}
        & \multicolumn{3}{c}{\textbf{RLBench}~\cite{rlbench}}
        & \multicolumn{3}{c}{\textbf{LIBERO}~\cite{libero}} \\
        \cmidrule(lr){3-5}
        \cmidrule(lr){6-8}
        &
        & SR$\uparrow$ (\%) & Speed$\uparrow$ & PR (\%)
        & SR$\uparrow$ (\%) & Speed$\uparrow$ & PR (\%) \\
        \midrule

        & Unpruned
        & \textbf{69.0} & 1.00x & 0/0
        & \textbf{88.0} & 1.00x & 0/0 \\
        \midrule

        \multirow{5}{*}{Att.}
        & FastV~\cite{fastv}
        & 65.7 & 1.71x & 75/75
        & 84.7 & 1.64x & 75/75 \\

        & SparseVLM~\cite{sparsevlm}
        & 66.5 & 1.66x & 75/75
        & 84.8 & 1.67x & 75/75 \\

        & SP-VLA~\cite{spvla}
        & 67.0 & 1.04x & 25/25
        & 86.1 & 1.09x & 25/25 \\

        & VLA-Pruner~\cite{vlapruner}
        & 67.5 & 1.61x & 75/75
        & 86.6 & 1.61x & 75/75 \\

        & Fast3D~\cite{fast3d}
        & 67.0 & 1.63x & 75/75
        & 85.9 & 1.66x & 75/75 \\
        \midrule

        \multirow{3}{*}{Fea.}
        & DivPrune~\cite{divprune}
        & 66.0 & 1.62x & 75/75
        & 84.7 & 1.69x & 75/75 \\

        & DART~\cite{dart}
        & 65.8 & 1.78x & 75/75
        & 84.7 & 1.78x & 75/75 \\

        & CDPruner~\cite{cdpruner}
        & 67.6 & 1.77x & 75/75
        & 86.7 & 1.71x & 75/75 \\
        \midrule

        Corr.
        & \textbf{\textit{2Dor3D}}
        & \textbf{68.4} & \textbf{1.86x} & 80/70
        & \textbf{87.5} & \textbf{1.91x} & 80/70 \\
        \bottomrule
        \bottomrule
    \end{tabular*}%
    }
    \vspace{-4mm}
\end{table}

For each method and backbone, each of five independent seeds covers 10 RLBench~\cite{rlbench} tasks (20 trials per task) and 40 LIBERO~\cite{kerv,libero} tasks (50 each). Success rate (SR) is the fraction of successes across 1,000 and 10,000 trials, respectively. End-to-end speedup compares closed-loop latency per step with the unpruned policy.

Our MLA~\cite{mla} and single-view FiS-VLA~\cite{fisvla} configurations each produce 256 image and 256 point tokens. A fixed 75/75 split retains 64 of each. Under the same cap $B=128$, \textit{2Dor3D} counts each shared token once (Eq.~\eqref{eq:unique_count}). Its 80/70 mean pruning rates summarize the allocation (51.2 image and 76.8 point tokens retained per step on average). The four-state model also chooses which patches retain appearance, geometry, both, or neither in each observation. SP-VLA uses 25/25.

We compare FastV~\cite{fastv}, SparseVLM~\cite{sparsevlm}, SP-VLA~\cite{spvla}, VLA-Pruner~\cite{vlapruner}, Fast3D~\cite{fast3d}, DivPrune~\cite{divprune}, DART~\cite{dart}, and CDPruner~\cite{cdpruner}. We use $K=4$, $L=2$, one NVIDIA A100-PCIe-40GB GPU, and an Intel Xeon Gold 6348 CPU.

\begin{figure}[b]
    \centering
    \vspace{-4mm}
    \includegraphics[width=3.3in]{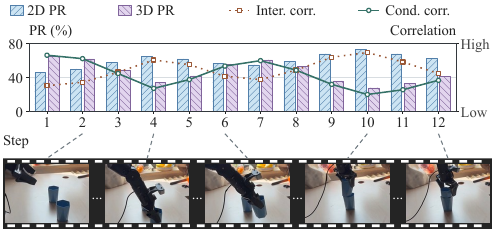}
    \vspace{-2mm}
    \caption{A Real-World Case Study of Token Pruning with \textit{2Dor3D}.}
    \label{fig:realworld}
\end{figure}

\subsection{\textbf{Main Results}}
\label{sec:experiments_2}

\subsubsection{\textbf{Simulator Results}}
Across RLBench and LIBERO (Tables~\ref{tab:main-mla} and~\ref{tab:main-fis}), \textit{2Dor3D} speeds up closed-loop inference by $2.06$--$2.10\times$ on MLA and $1.86$--$1.91\times$ on FiS-VLA. Observed mean SR is 0.5--0.6 percentage points below the unpruned policies.

Relative to CDPruner, the most accurate pruning baseline, \textit{2Dor3D} runs at 4.8\%--10.7\% lower latency with 0.6--0.8 percentage points higher mean SR. Its mean SR exceeds VLA-Pruner's by 0.8--0.9 points. At the same 128-token budget as DART, latency falls by 4.3\%--10.0\% and mean SR rises by 1.6--2.8 points.

\begin{table}[t]
    \centering
    \caption{Real-World Evaluation Results}
    \vspace{-2mm}
    \label{tab:real-world}
    \footnotesize
    \setlength{\tabcolsep}{2pt}
    \renewcommand{\arraystretch}{0.98}
    \setlength{\aboverulesep}{0.2ex}
    \setlength{\belowrulesep}{0.2ex}
    \setlength{\cmidrulesep}{0.15ex}

    \begin{tabular*}{\columnwidth}{
        @{\extracolsep{\fill}}c*{8}{c}@{}
    }
        \toprule
        \toprule
        \multirow{2}{*}[-0.5ex]{Method}
        & \multicolumn{4}{c}{Short-horizon}
        & \multicolumn{2}{c}{Multi-stage}
        & SR$\uparrow$
        & Task Spd.$\uparrow$ \\
        \cmidrule(lr){2-5}
        \cmidrule(lr){6-7}
        & T1 & T2 & T3 & T4 & T5 & T6
        & (\%) & (x) \\
        \midrule

        \multicolumn{9}{c}{\textbf{MLA}~\cite{mla}} \\
        \midrule

        Unpruned
        & 19/20 & 16/20 & 17/20 & 18/20 & 15/20 & 14/20
        & \textbf{82.5} & 1.00 \\

        Fast3D~\cite{fast3d}
        & 18/20 & 14/20 & 16/20 & 17/20 & 12/20 & 12/20
        & 74.2 & 1.35 \\

        VLA-Pruner~\cite{vlapruner}
        & 18/20 & 15/20 & 16/20 & 18/20 & 13/20 & 13/20
        & 77.5 & 1.32 \\

        \textbf{\textit{2Dor3D}}
        & 19/20 & 16/20 & 17/20 & 18/20 & 14/20 & 14/20
        & \textbf{81.7} & \textbf{1.45} \\
        \midrule

        \multicolumn{9}{c}{\textbf{FiS-VLA}~\cite{fisvla}} \\
        \midrule

        Unpruned
        & 18/20 & 15/20 & 16/20 & 17/20 & 13/20 & 12/20
        & \textbf{75.8} & 1.00 \\

        Fast3D~\cite{fast3d}
        & 17/20 & 13/20 & 14/20 & 16/20 & 10/20 & 10/20
        & 66.7 & 1.21 \\

        VLA-Pruner~\cite{vlapruner}
        & 18/20 & 14/20 & 15/20 & 16/20 & 11/20 & 11/20
        & 70.8 & 1.21 \\

        \textbf{\textit{2Dor3D}}
        & 18/20 & 15/20 & 16/20 & 17/20 & 12/20 & 12/20
        & \textbf{75.0} & \textbf{1.26} \\
        \bottomrule
        \bottomrule
    \end{tabular*}
    \vspace{-4mm}
\end{table}

\begin{table}[b]
    \centering
    \vspace{-4mm}
    \caption{Progressive Ablation on RLBench~\cite{rlbench} Simulation Benchmark}
    \label{tab:ablation-progressive}
    \vspace{-2mm}
    \footnotesize
    \setlength{\tabcolsep}{2pt}
    \renewcommand{\arraystretch}{0.98}
    \setlength{\aboverulesep}{0.2ex}
    \setlength{\belowrulesep}{0.2ex}
    \setlength{\cmidrulesep}{0.15ex}

    \begin{tabular*}{\columnwidth}{
        @{\extracolsep{\fill}}c*{4}{c}@{}
    }
        \toprule
        \toprule
        \multirow{2}{*}{Configuration}
        & \multicolumn{2}{c}{MLA~\cite{mla}}
        & \multicolumn{2}{c}{FiS-VLA~\cite{fisvla}} \\
        \cmidrule(lr){2-3}
        \cmidrule(lr){4-5}
        & SR$\uparrow$ (\%) & Speedup$\uparrow$ (x)
        & SR$\uparrow$ (\%) & Speedup$\uparrow$ (x) \\
        \midrule

        Unpruned
        & 81.0 & 1.00
        & 69.0 & 1.00 \\
        \midrule

        Base
        & 75.1 & \textbf{2.10}
        & 62.7 & \textbf{1.98} \\

        + Candidate Modeling
        & 78.2 & 2.09
        & 65.9 & 1.95 \\

        + Guided Exploration
        & 79.7 & 2.07
        & 67.4 & 1.90 \\

        \textbf{+ Coupling Optimization}
        & \textbf{80.5} & 2.06
        & \textbf{68.4} & 1.86 \\
        \bottomrule
        \bottomrule
    \end{tabular*}
\end{table}

\subsubsection{\textbf{Real-World Evaluation}}
The six DyQ-VLA~\cite{dyq} tasks are placing a banana on a plate (T1), nesting cups (T2), inserting a banana into a cup (T3), placing an apple on a plate (T4), packing an orange (T5), and placing fruit in order (T6). For each backbone, all methods share a checkpoint fine-tuned on human demonstrations and 20 matched initial configurations per task.

Across 120 matched trials per backbone (Table~\ref{tab:real-world}), \textit{2Dor3D} matches the unpruned policy on T1--T4 and T6, with one fewer success on T5. SR reaches 81.7\% on MLA and 75.0\% on FiS-VLA, 4.2 points above VLA-Pruner on each. Including robot motion, task completion speeds up by $1.45\times$ and $1.26\times$, versus $1.35\times$ and $1.21\times$ for the fastest baselines.

Fig.~\ref{fig:realworld} shows how the modality allocation changes during cup nesting (T2). At step~2, the policy prunes a larger fraction of 3D tokens than of 2D tokens. By step~10, this ordering reverses as the gripper aligns the cups.

\subsection{\textbf{Ablation Study}}
\label{sec:experiments_3}
Table~\ref{tab:ablation-progressive} starts from a fixed-budget greedy selector with independent modality costs and no compensation. State scoring and residual compensation together raise SR by 3.1 and 3.2 percentage points on MLA and FiS-VLA, respectively. Neighborhood-guided search adds 1.5 points on both backbones. Local exchanges with coupling-aware scoring add a further 0.8 and 1.0 points, bringing SR to 80.5\% and 68.4\%, respectively. At the same token budget, the full method gains 5.4 and 5.7 percentage points in SR over the greedy baseline on MLA and FiS-VLA, respectively. 
With state scoring and residual compensation already enabled, the two search stages contribute 43\% of the total SR gain on MLA and 44\% on FiS-VLA in this ablation sequence.
Speedup falls only from $2.10\times$ to $2.06\times$ on MLA and from $1.98\times$ to $1.86\times$ on FiS-VLA.

\subsection{\textbf{Discussion}}
\label{sec:experiments_4}

\subsubsection{\textbf{Hyperparameters}}
\begin{figure}[b]
    \centering
    \vspace{-4mm}
    \includegraphics[width=3.3in]{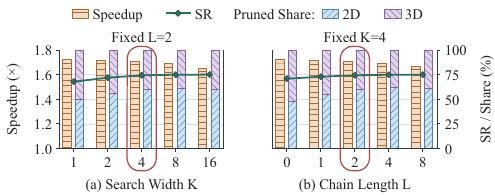}
    \vspace{-2mm}
    \caption{Sensitivity to Search Width $K$ and Chain Length $L$.}
    \label{fig:hp}
\end{figure}

Under a fixed token budget, Fig.~\ref{fig:hp} plots speedup, SR, and each modality's share of the pruned tokens. Panel~(a) varies search width $K$ at $L=2$, and panel~(b) varies chain length $L$ at $K=4$.

\begin{figure}[b]
    \centering
    \vspace{-4mm}
    \includegraphics[width=3.3in]{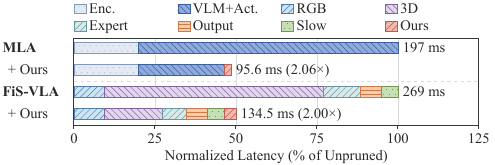}
    \vspace{-2mm}
    \caption{Measured Single-Step Model Latency by Stage on RLBench.}
    \label{fig:breakdown}
\end{figure}

Row-major traversal avoids sorting patches for each observation. At $K=4$, the search retains up to four partial allocations. At $L=2$, up to two refinement moves can release tokens from earlier patches for later ones. Beyond these settings, success rate changes little while speedup decreases.

\subsubsection{\textbf{Latency Breakdown}}
Fig.~\ref{fig:breakdown} reports single-step model latency on RLBench. On MLA, tokenization stays at 39\,ms. Shorter sequences save 105.4\,ms in the backbone and action stages (158 to 52.6\,ms). Pruning costs 4\,ms, leaving a net saving of 101.4\,ms ($2.06\times$). On FiS-VLA, encoder compaction shortens the sequence for downstream blocks and the fast expert. The 3D stage drops from 182 to 48.5\,ms, RGB stays at 25\,ms, and total model latency falls from 269 to 134.5\,ms ($2.00\times$).

Correlation estimation, search, refinement, compensation, and compaction together take 4.2\% of pruned MLA latency and 7.4\% (10\,ms) on FiS-VLA. The FiS-VLA breakdown amortizes slow-system updates across steps. Closed-loop timing (Table~\ref{tab:main-fis}) includes depth-to-point-cloud preprocessing and actual slow-system calls at episode boundaries, yielding $1.86\times$ on RLBench and $1.91\times$ on LIBERO.

\section{\textbf{Conclusion}}
\label{sec:conclusion}

We present \textit{2Dor3D}, a framework that uses cross-modal correlations to jointly prune image and point cloud tokens in 3D VLAs.
It transfers cross-modal residuals to retained tokens, guides the search with neighborhood correlations, and updates pruning costs across coupled patches.
Later layers thus process fewer tokens while preserving visual and geometric cues for action prediction.

\balance
\clearpage
\bibliographystyle{IEEEtran}
\bibliography{ref/references}

@article{rt2,
  title={Rt-2: Vision-language-action models transfer web knowledge to robotic control},
  author={Brohan, Anthony and Brown, Noah and Carbajal, Justice and Chebotar, Yevgen and Chen, Xi and Choromanski, Krzysztof and Ding, Tianli and Driess, Danny and Dubey, Avinava and Finn, Chelsea and others},
  journal={arXiv preprint arXiv:2307.15818},
  year={2023}
}

@article{openvla,
  title={Openvla: An open-source vision-language-action model},
  author={Kim, Moo Jin and Pertsch, Karl and Karamcheti, Siddharth and Xiao, Ted and Balakrishna, Ashwin and Nair, Suraj and Rafailov, Rafael and Foster, Ethan and Lam, Grace and Sanketi, Pannag and others},
  journal={arXiv preprint arXiv:2406.09246},
  year={2024}
}

@inproceedings{mla,
  title={Mla: A multisensory language-action model for multimodal understanding and forecasting in robotic manipulation},
  author={Liu, Zhuoyang and Liu, Jiaming and Xu, Jiadong and Han, Nuowei and Gu, Chenyang and Chen, Hao and Zhou, Kaichen and Zhang, Renrui and Hsieh, Kai Chin and Wu, Kun and others},
  booktitle={2026 IEEE International Conference on Robotics and Automation (ICRA)},
  pages={5510--5517},
  year={2026},
  organization={IEEE}
}

@article{pointvla,
  title={Pointvla: Injecting the 3d world into vision-language-action models},
  author={Li, Chengmeng and Wen, Junjie and Peng, Yaxin and Peng, Yan and Zhu, Yichen},
  journal={IEEE Robotics and Automation Letters},
  volume={11},
  number={3},
  pages={2506--2513},
  year={2026},
  publisher={IEEE}
}

@inproceedings{3dsvla,
  title={3ds-vla: A 3d spatial-aware vision language action model for robust multi-task manipulation},
  author={Li, Xiaoqi and Heng, Liang and Liu, Jiaming and Shen, Yan and Gu, Chenyang and Liu, Zhuoyang and Chen, Hao and Han, Nuowei and Zhang, Renrui and Tang, Hao and others},
  booktitle={9th Annual Conference on Robot Learning},
  year={2025}
}

@article{spatialvla,
  title={Spatialvla: Exploring spatial representations for visual-language-action model},
  author={Qu, Delin and Song, Haoming and Chen, Qizhi and Yao, Yuanqi and Ye, Xinyi and Ding, Yan and Wang, Zhigang and Gu, JiaYuan and Zhao, Bin and Wang, Dong and others},
  journal={arXiv preprint arXiv:2501.15830},
  year={2025}
}

@article{3dvla,
  title={3d-vla: A 3d vision-language-action generative world model},
  author={Zhen, Haoyu and Qiu, Xiaowen and Chen, Peihao and Yang, Jincheng and Yan, Xin and Du, Yilun and Hong, Yining and Gan, Chuang},
  journal={arXiv preprint arXiv:2403.09631},
  year={2024}
}

@inproceedings{fast3d,
  title={Fast3d: Accelerating 3d multi-modal large language models for efficient 3d scene understanding},
  author={Huang, Wencan and Liu, Daizong and Hu, Wei},
  booktitle={Proceedings of the 33rd ACM International Conference on Multimedia},
  pages={10935--10944},
  year={2025}
}

@article{geo3dpruner,
  title={Geometry-Guided 3D Visual Token Pruning for Video-Language Models},
  author={Li, Han and Huang, Zehao and Fu, Jiahui and Wang, Naiyan and Liu, Si},
  journal={arXiv preprint arXiv:2604.18260},
  year={2026}
}

@inproceedings{tokenfusion,
  title={Multimodal token fusion for vision transformers},
  author={Wang, Yikai and Chen, Xinghao and Cao, Lele and Huang, Wenbing and Sun, Fuchun and Wang, Yunhe},
  booktitle={2022 IEEE/CVF Conference on Computer Vision and Pattern Recognition (CVPR)},
  pages={12176--12185},
  year={2022},
  organization={IEEE}
}

@article{efficientvla,
  title={Efficientvla: Training-free acceleration and compression for vision-language-action models},
  author={Yang, Yantai and Wang, Yuhao and Wen, Zichen and Zhongwei, Luo and Zou, Chang and Zhang, Zhipeng and Wen, Chuan and Zhang, Linfeng},
  journal={Advances in Neural Information Processing Systems},
  volume={38},
  pages={40891--40914},
  year={2026}
}

@article{specprune-vla,
  title={Specprune-vla: Accelerating vision-language-action models via action-aware self-speculative pruning},
  author={Wang, Hanzhen and Xu, Jiaming and Xiang, Yushun and Pan, Jiayi and Zhou, Yongkang and Li, Yong-Lu and Dai, Guohao},
  journal={arXiv preprint arXiv:2509.05614},
  year={2025}
}

@article{leo,
  title={An embodied generalist agent in 3d world},
  author={Huang, Jiangyong and Yong, Silong and Ma, Xiaojian and Linghu, Xiongkun and Li, Puhao and Wang, Yan and Li, Qing and Zhu, Song-Chun and Jia, Baoxiong and Huang, Siyuan},
  journal={arXiv preprint arXiv:2311.12871},
  year={2023}
}

@article{fisvla,
  title={Fast-in-Slow: a dual-system VLA model unifying fast manipulation within slow reasoning},
  author={Chen, Hao and Liu, Jiaming and Gu, Chenyang and Liu, Zhuoyang and Zhang, Renrui and Li, Xiaoqi and He, Xiao and Guo, Yandong and Fu, Chi-Wing and Zhang, Shanghang and others},
  journal={Advances in Neural Information Processing Systems},
  volume={38},
  pages={98049--98083},
  year={2026}
}

@article{rlbench,
  title={Rlbench: The robot learning benchmark \& learning environment},
  author={James, Stephen and Ma, Zicong and Arrojo, David Rovick and Davison, Andrew J},
  journal={IEEE Robotics and Automation Letters},
  volume={5},
  number={2},
  pages={3019--3026},
  year={2020},
  publisher={IEEE}
}

@inproceedings{spvla,
  title={Sp-vla: A joint model scheduling and token pruning approach for vla model acceleration},
  author={Li, Ye and Meng, Yuan and Sun, Zewen and Ji, Kangye and Tang, Chen and Fan, Jiajun and Ma, Xinzhu and Xia, Shu-Tao and Wang, Zhi and Zhu, Wenwu},
  booktitle={The Fourteenth International Conference on Learning Representations},
  year={2025}
}

@article{vlapruner,
  title={Vla-pruner: Temporal-aware dual-level visual token pruning for efficient vision-language-action inference},
  author={Liu, Ziyan and Chen, Yeqiu and Cai, Hongyi and Lin, Tao and Yang, Shuo and Liu, Zheng and Zhao, Bo},
  journal={arXiv e-prints},
  pages={arXiv--2511},
  year={2025}
}

@article{libero,
  title={Libero: Benchmarking knowledge transfer for lifelong robot learning},
  author={Liu, Bo and Zhu, Yifeng and Gao, Chongkai and Feng, Yihao and Liu, Qiang and Zhu, Yuke and Stone, Peter},
  journal={Advances in Neural Information Processing Systems},
  volume={36},
  pages={44776--44791},
  year={2023}
}

@inproceedings{fastv,
  title={An image is worth 1/2 tokens after layer 2: Plug-and-play inference acceleration for large vision-language models},
  author={Chen, Liang and Zhao, Haozhe and Liu, Tianyu and Bai, Shuai and Lin, Junyang and Zhou, Chang and Chang, Baobao},
  booktitle={European Conference on Computer Vision},
  pages={19--35},
  year={2024},
  organization={Springer}
}

@article{sparsevlm,
  title={Sparsevlm: Visual token sparsification for efficient vision-language model inference},
  author={Zhang, Yuan and Fan, Chun-Kai and Ma, Junpeng and Zheng, Wenzhao and Huang, Tao and Cheng, Kuan and Gudovskiy, Denis and Okuno, Tomoyuki and Nakata, Yohei and Keutzer, Kurt and others},
  journal={arXiv preprint arXiv:2410.04417},
  year={2024}
}

@inproceedings{divprune,
  title={Divprune: Diversity-based visual token pruning for large multimodal models},
  author={Alvar, Saeed Ranjbar and Singh, Gursimran and Akbari, Mohammad and Zhang, Yong},
  booktitle={2025 IEEE/CVF Conference on Computer Vision and Pattern Recognition (CVPR)},
  pages={9392--9401},
  year={2025},
  organization={IEEE}
}

@inproceedings{dart,
  title={Stop looking for “important tokens” in multimodal language models: Duplication matters more},
  author={Wen, Zichen and Gao, Yifeng and Wang, Shaobo and Zhang, Junyuan and Zhang, Qintong and Li, Weijia and He, Conghui and Zhang, Linfeng},
  booktitle={Proceedings of the 2025 Conference on Empirical Methods in Natural Language Processing},
  pages={9972--9991},
  year={2025}
}

@article{cdpruner,
  title={Beyond attention or similarity: Maximizing conditional diversity for token pruning in mllms},
  author={Zhang, Qizhe and Liu, Mengzhen and Li, Lichen and Lu, Ming and Zhang, Yuan and Pan, Junwen and She, Qi and Zhang, Shanghang},
  journal={Advances in Neural Information Processing Systems},
  volume={38},
  pages={25438--25468},
  year={2026}
}

@article{dyq,
  title={Dyq-vla: Temporal-dynamic-aware quantization for embodied vision-language-action models},
  author={Zheng, Zihao and Cao, Hangyu and Tao, Chibang and Mao, Zhihao and Zhang, Lingyue and Li, Maoliang and Chen, Jiayu and Zou, Hailong and Liu, Xuanzhe and Cao, Donggang and others},
  journal={arXiv preprint arXiv:2603.07904},
  year={2026}
}

@inproceedings{kerv,
  title={Kerv: Kinematic-rectified speculative decoding for embodied vla models},
  author={Zheng, Zihao and Mao, Zhihao and Li, Maoliang and Chen, Jiayu and Sun, Xinhao and Zhang, Zhaobo and Cao, Donggang and Mei, Hong and Chen, Xiang},
  booktitle={Proceedings of the 63rd ACM/IEEE Design Automation Conference},
  pages={1--7},
  year={2026}
}

@article{rt1,
  title={Rt-1: Robotics transformer for real-world control at scale},
  author={Brohan, Anthony and Brown, Noah and Carbajal, Justice and Chebotar, Yevgen and Dabis, Joseph and Finn, Chelsea and Gopalakrishnan, Keerthana and Hausman, Karol and Herzog, Alex and Hsu, Jasmine and others},
  journal={arXiv preprint arXiv:2212.06817},
  year={2022}
}

@article{geovla,
  title={Geovla: Empowering 3d representations in vision-language-action models},
  author={Sun, Lin and Xie, Bin and Liu, Yingfei and Shi, Hao and Wang, Tiancai and Cao, Jiale},
  journal={arXiv preprint arXiv:2508.09071},
  year={2025}
}

@inproceedings{imvotenet,
  title={Imvotenet: Boosting 3d object detection in point clouds with image votes},
  author={Qi, Charles R and Chen, Xinlei and Litany, Or and Guibas, Leonidas J},
  booktitle={2020 IEEE/CVF conference on computer vision and pattern recognition (CVPR)},
  pages={4403--4412},
  year={2020},
  organization={IEEE}
}

@article{dynamicvit,
  title={Dynamicvit: Efficient vision transformers with dynamic token sparsification},
  author={Rao, Yongming and Zhao, Wenliang and Liu, Benlin and Lu, Jiwen and Zhou, Jie and Hsieh, Cho-Jui},
  journal={Advances in neural information processing systems},
  volume={34},
  pages={13937--13949},
  year={2021}
}

@article{diffactor,
  title={3d diffuser actor: Policy diffusion with 3d scene representations},
  author={Ke, Tsung-Wei and Gkanatsios, Nikolaos and Fragkiadaki, Katerina},
  journal={arXiv preprint arXiv:2402.10885},
  year={2024}
}

@inproceedings{lightvla,
  title={The better you learn, the smarter you prune: Towards efficient vision-language-action models via differentiable token pruning},
  author={Jiang, Titong and Jiang, Xuefeng and Ma, Yuan and Wen, Xin and Li, Bailin and Zhan, Kun and Jia, Peng and Liu, Yahui and Sun, Sheng and Lang, Xianpeng},
  booktitle={2026 IEEE International Conference on Robotics and Automation (ICRA)},
  pages={20645--20652},
  year={2026},
  organization={IEEE}
}

@article{iaprune,
  title={Training-Free Interaction-Aligned Visual Token Pruning for Efficient Embodied Manipulation},
  author={Cheng, Jintao and Li, Weibin and Wang, Haozhe and Wang, Gang and Zhang, Yipu and Tang, Xiaoyu and Wu, Jin and Chen, Xieyuanli and Liu, Yunhui and Zhang, Wei},
  journal={arXiv preprint arXiv:2603.22991},
  year={2026}
}

@inproceedings{flashattention2,
  title={Flashattention-2: Faster attention with better parallelism and work partitioning},
  author={Dao, Tri},
  booktitle={International Conference on Learning Representations},
  volume={2024},
  pages={35549--35562},
  year={2024}
}

\end{document}